\documentstyle[psfig,epsfig]{mn}
\input epsf
 
\begin{document}
 
\newcommand{\kms}{\,{\rm km}\,{\rm s}^{-1}}
\newcommand{\mum}{$\,\mu$m}
\newcommand{\eg}{e.g., }
\newcommand{\ie}{i.e., }
\newcommand{\cf}{cf.,}
\newcommand{\etal}{et~al.\ }
\newcommand{\msun}{\mbox{$M_\odot$}}
\newcommand{\ha}{{\it H$\alpha$}~}
\newcommand{\hb}{{\it H$\beta$}~}
\newcommand{\xx}{$\times$}
\newcommand{\magsq}{magnitudes arcsec${^-2}$~}
 
\def\gs{\mathrel{\raise0.35ex\hbox{$\scriptstyle >$}\kern-0.6em
\lower0.40ex\hbox{{$\scriptstyle \sim$}}}}
\def\ls{\mathrel{\raise0.35ex\hbox{$\scriptstyle <$}\kern-0.6em
\lower0.40ex\hbox{{$\scriptstyle \sim$}}}}
 
\journal{Preprint UBC-AST-98-03, astro-ph/9912032}
\title[Adaptive optics imaging of NGC\,3227]
{Near-IR adaptive optics imaging of nuclear spiral structure in
the Seyfert galaxy, NGC\,3227}

\author[S.\,C.\ Chapman et al.]
       {Scott C.\ Chapman$^{\! 1,2,4}$, Simon L.\ Morris$^{\! 3,4}$,
	and Gordon A. H.\ Walker$^{\! 1}$
        \vspace*{1mm}\\
[1] Department\ of Physics \& Astronomy,
        University of British Columbia,
        Vancouver, B.C.~V6T 1Z4,~~Canada\\
[2] Present address: Observatories of the Carnegie Institution of
        Washington, Pasadena, CA 91101,~~U.S.A.\\
[3] Dominion Astrophysics Observatory, Victoria, B.C.,~~Canada V8X 4M6\\
[4] Guest observer, Canada France Hawaii Telescope\\        }

%\date{\fbox{\sc Submitted Version}}
\date{Accepted ... ;
      Received ... ;
      in original form ...}
\pubyear{1998}
\pagerange{000--000}
\maketitle
 
\begin{abstract}
We present high spatial resolution, near-IR images in $J$,$H$, and $K$ of the
nucleus of NGC~3227, obtained with the Adaptive Optics bonnette on
CFHT. The $\sim0.15$\arcsec\ (17pc) resolution allows structures to be 
probed in the core region. 
Dust obscuration becomes significantly less pronounced at longer wavelengths,
revealing the true geometry of the core region.
We are able to identify two main features in our maps: 
{\it (i)}
a spiraling association of knots with a counterpart in an HST F606W image;
{\it (ii)}
a smaller scale annulus, orthogonal to the spiral of knots.
These features may provide a means to transport material inwards to fuel the
active nucleus.
%The observed  structures may help to elucidate
%fueling mechanisms for the central engine, as well as providing insight into
%the standard unification picture.
\end{abstract}

 
%%%%%%%%%%%%%%%%%%%%%%%%%%%%%%%%%%%%%%%%%%%%%%%%%%%%%%%%%%%%%%%%%%%%%%%%%
 
\section{Introduction}

It is generally accepted that pronounced activity in
galaxies hosting Active Galactic Nuclei (AGN) results from accretion onto
a supermassive black hole.  
%This paradigm has led to a plethora of
%research into AGN, of which 
However, the problems of overcoming the angular momentum
barrier to fuel the nucleus \cite{shlosman89} and also the
unification of the AGN types \cite{ant93} 
continue to be vexing and controversial.
Near-IR imaging has proven to be  a powerful means to
study these AGN problems, since the dust extinction is reduced,
as is the contrast between the central AGN and the underlying stellar population
(McLeod et al. 1995).
A fresh perspective can be gained on the core regions of AGNs through the
high resolution
images now possible thanks to adaptive optics (AO).

Here, we present observations of  NGC~3227  obtained  with {\it PUEO}, the AO
system  recently commissioned on the 3.6 m Canada-France-Hawaii
Telescope \cite{rigaut}. NGC~3227 is an SABa galaxy, interacting with its dwarf elliptical neighbor,
NGC~3226. It has been much studied in recent years as it contains many of the
elements thought to be related to the formation and evolution of active
nuclei: emission line regions excited by both starburst and AGN continuum,
strong interaction, and a stellar bar (Gonzales Delgado \& Perez 1997, Arribas \& Mediavilla 1994).

\section{Observations}
The images were obtained at the CFHT in March, 1997, using the MONICA
near-IR camera (Nadeau et al. 1994) mounted at the f/20 focus of the
Adaptive Optics Bonnette (AOB). The
detector is a Rockwell NICMOS3 array with 256\xx\,256 pixels and a 
0.034\arcsec\ per pixel
scale.
The CFHT AOB is based on curvature wavefront sensing (Roddier 1991),
and uses a  19 zone bimorph
mirror to correct the wavefront distortions \cite{rigaut}.
As the field size is small with MONICA (9\arcsec\xx\,9\arcsec\,), blank sky images were taken intermitantly
between science frames. On-source images were taken in a mosaic of 4 positions,
alternately putting the galaxy core in each of the four quadrants of the
array.
Flux and PSF calibrations were performed using the UKIRT standard stars
fs13 and fs25. Flat-field images were taken on the dome with the lamps turned
on and off to account for the thermal glow of the telescope. The nucleus
of the galaxy itself was used as the guiding source for the AO system,
roughly a 15th magnitude point source. The natural seeing averaged 0.6\arcsec\,-0.8\arcsec\
throughout the observations resulting in relatively high strehl ratios in
all bands, and FWHM of 0.14\arcsec\,, 0.17\arcsec\,, 0.22\arcsec\ at $K$, $H$, $J$ bands respectively.
At a distance of
15\,Mpc for NGC\,3227, 1\arcsec\,=76\,pc using H$_0=50$~km\,s$^{-1}$Mpc$^{-1}$.

Image processing proceeded as follows:  {\it i)} bad pixel correction; {\it ii)}
sky  subtraction,  using a  median-averaged  sky estimate; {\it iii)} flat-field
correction;  {\it  iv)}   re-centering  of  the  different   exposures   through
cross-correlation  techniques;  {\it v)}  adjustment  of the sky level among the
overlapping  regions to produce a homogeneous  background; {\it vi)} co-addition
of the  overlapping  regions,  rejecting  deviant  pixels  (clipped  mean).  The
resulting  images  were then  deconvolved  using the  classical  Lucy-Richardson
algorithm (25 iterations) 
with an input PSF reconstructed from the AO modal control commands obtained
during the actual observations \cite{veran97}.
%while following the recipe recently proposed by Magain et al. (1997) to constrain the final PSF.

\subsection{HST archive image}
An HST WFPC2 image was retrieved from the archive for NGC3227 in the F606W filter, corresponding 
roughly to Johnson $V$ and $R$ bands. The strong \ha\ line
is contained in this filter, and can contribute as much as several percent to 
integrated flux in this band for some observed morphologies.
The core of the galaxy lies on the PC chip with 0\farcs04 pixels. The image was rotated and rebinned to match
the pixel sampling and resolution of our AOB images.
The core of the galaxy is actually saturated and CCD bleeding is seen lying along the NE-SW axis.

\section{Results}

\subsection{Core structures}
The CFHT $J$, $H$, $K$ and HST F606W (hereafter $V$) -band images are presented in 
figure 1,  on a magnitude (log) scale. 
%The dynamic range of AOB is large (typically 1.3\xx\,10$^4$ at
%K) allows  significant details to be seen at all flux levels.
Subtraction of a smooth model galaxy (usually called {\it unsharp masking} -- 
see below) reveals that this region is punctuated with bright
knotty structures within a region 3\arcsec\xx\,2\arcsec\
(figure 2). The knots appear to trace out several ring patterns, and
are suggestive of a one-armed spiral.
If the assembly is interpreted as a spiral, the winding sense is 
counterclockwise, in agreement with the large scale spiral arms.
The colours of the knots are consistent with stellar and nuclear continuum
contributions (section 3.2).
We explored several methods of removing the low frequency galactic component, including various smoothing filters, a one-dimensional elliptical isophote model, and a multi-component (bulge, disk, point-source) elliptical isophote model.
All methods consistently unveil the spiraling structure, with knots coincident in all of $J$, $H$, $K$- and HST $V$-bands.
However, in the central 0.5 arcsec of the galaxy,
subtracting
isophotal fitting models (unsharp masking)
reveals prominent artifacts which obscure structural details.
as described in Chapman et al. (1999a, 1998).

We note that there has been some controversy over artifacts at the
core of AGN galaxies imaged with AO, 
which imitate expected
AGN structures (spiral arms, tori, bars, etc.).
Firstly,
all structures treated as physical in this paper 
are either further from the nucleus,
or else they are distinctly different and
more extended than the
artifacts cataloged in Chapman et al. (1998, 1999a). 
%, as with the case of the blue nuclear disk.
Secondly, we have compared the general morphology with HST-NICMOS images 
(Alonso-Herrero, private communication), 
%Check the CADC: mulchaey's image may be available now
which shows similar features, although at a lower spatial resolution at K-band
and with 0.1\arcsec\ pixels compared with our 0.034\arcsec\ pixels.

%
% FIGURE 1
%
\begin{figure*}  
\begin{minipage}{170mm}
%(a) \hskip 62.5mm (b)
\begin{center}
%\vskip -5mm
\epsfig{file=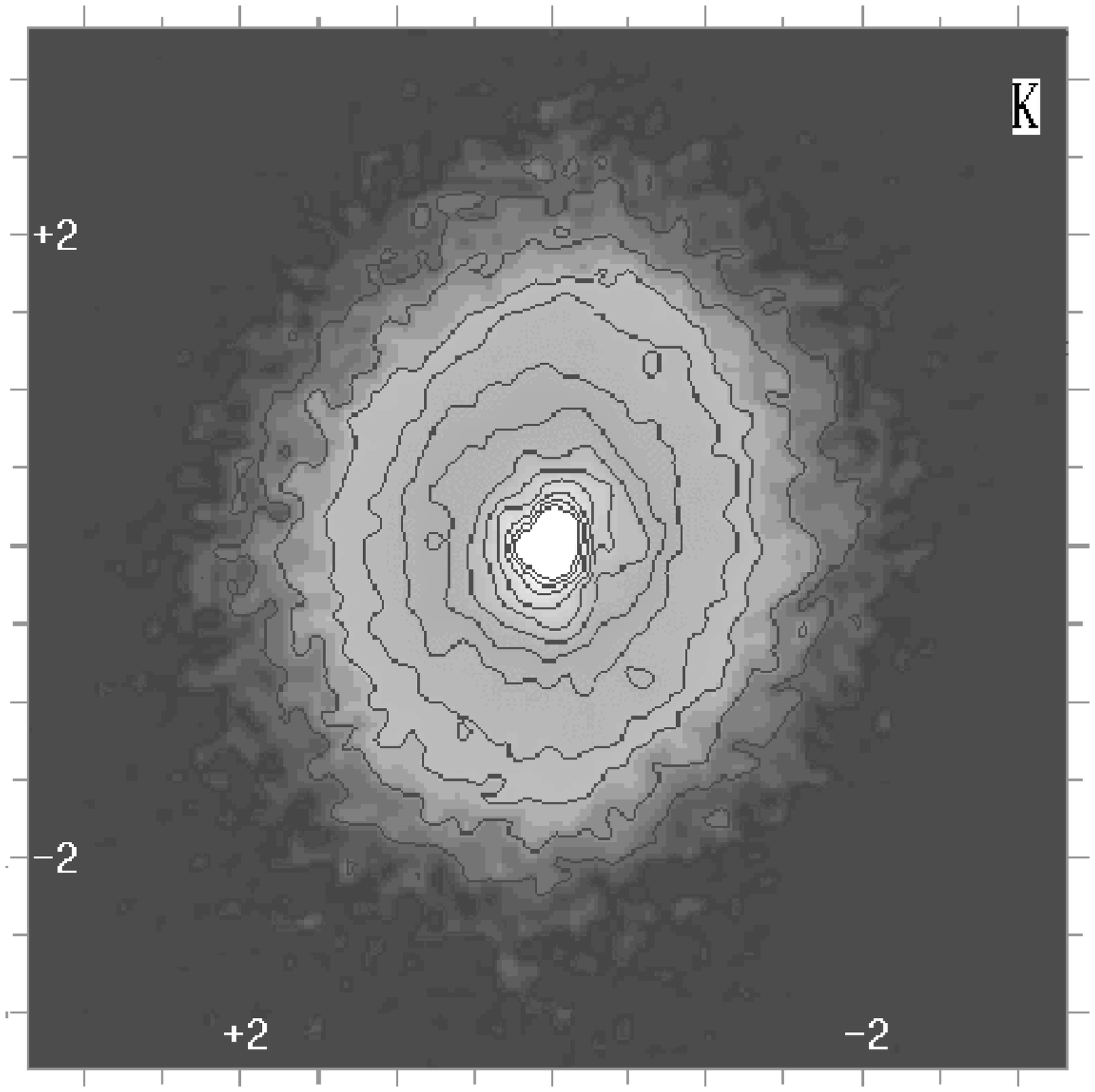, width=8.25cm, angle=0}
\epsfig{file=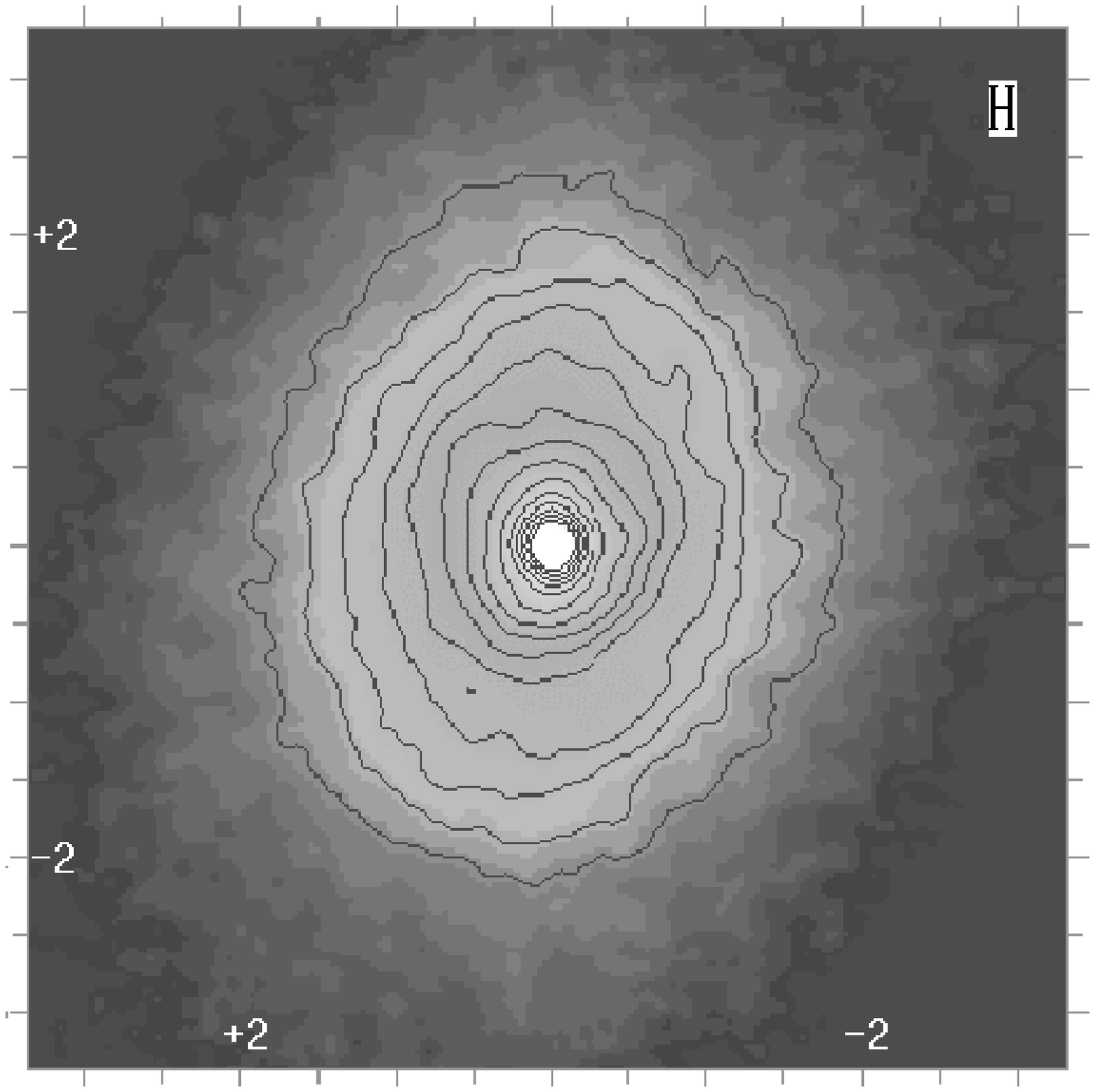, width=8.25cm, angle=0}
\epsfig{file=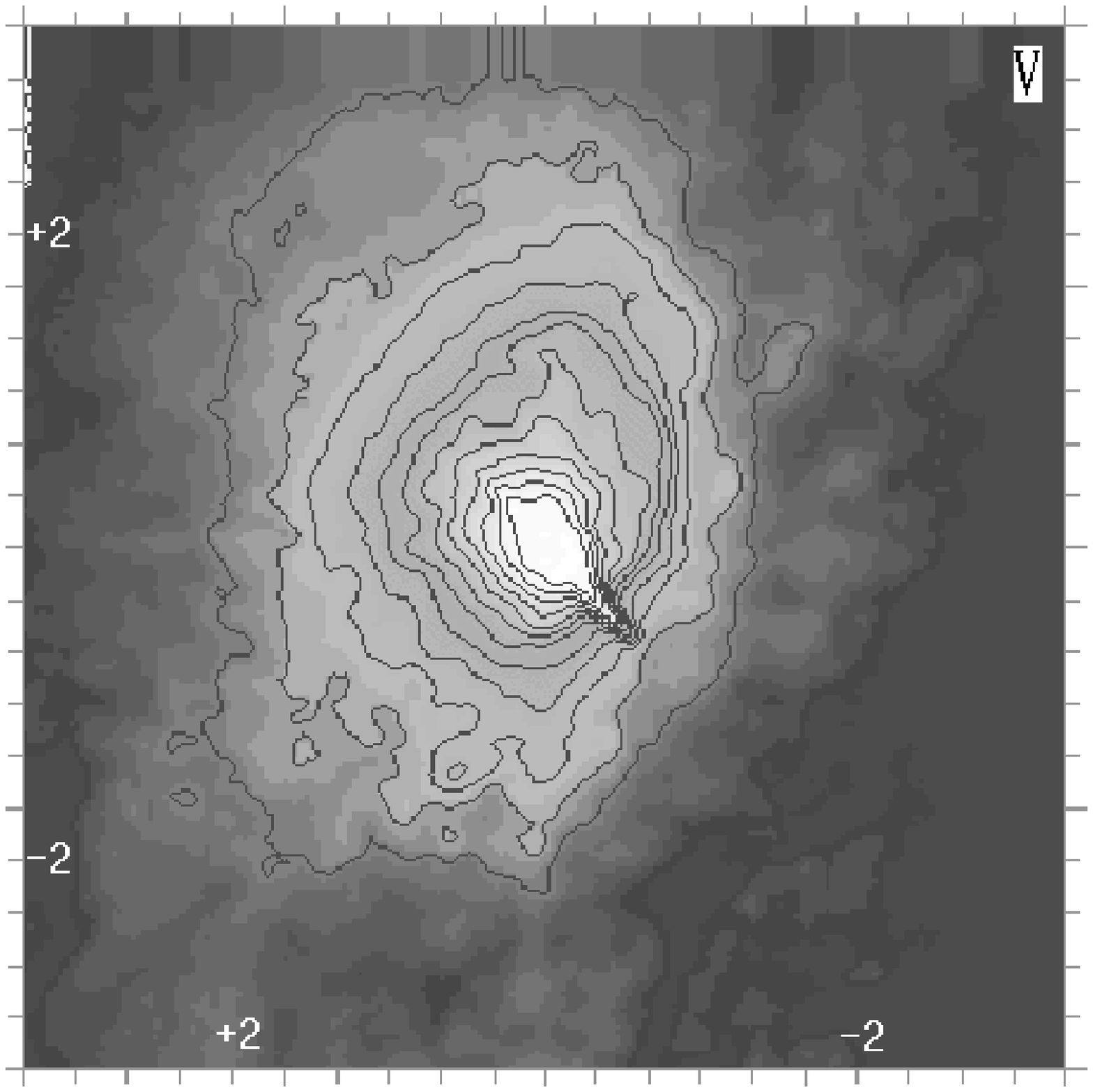, width=8.25cm, angle=0}
\epsfig{file=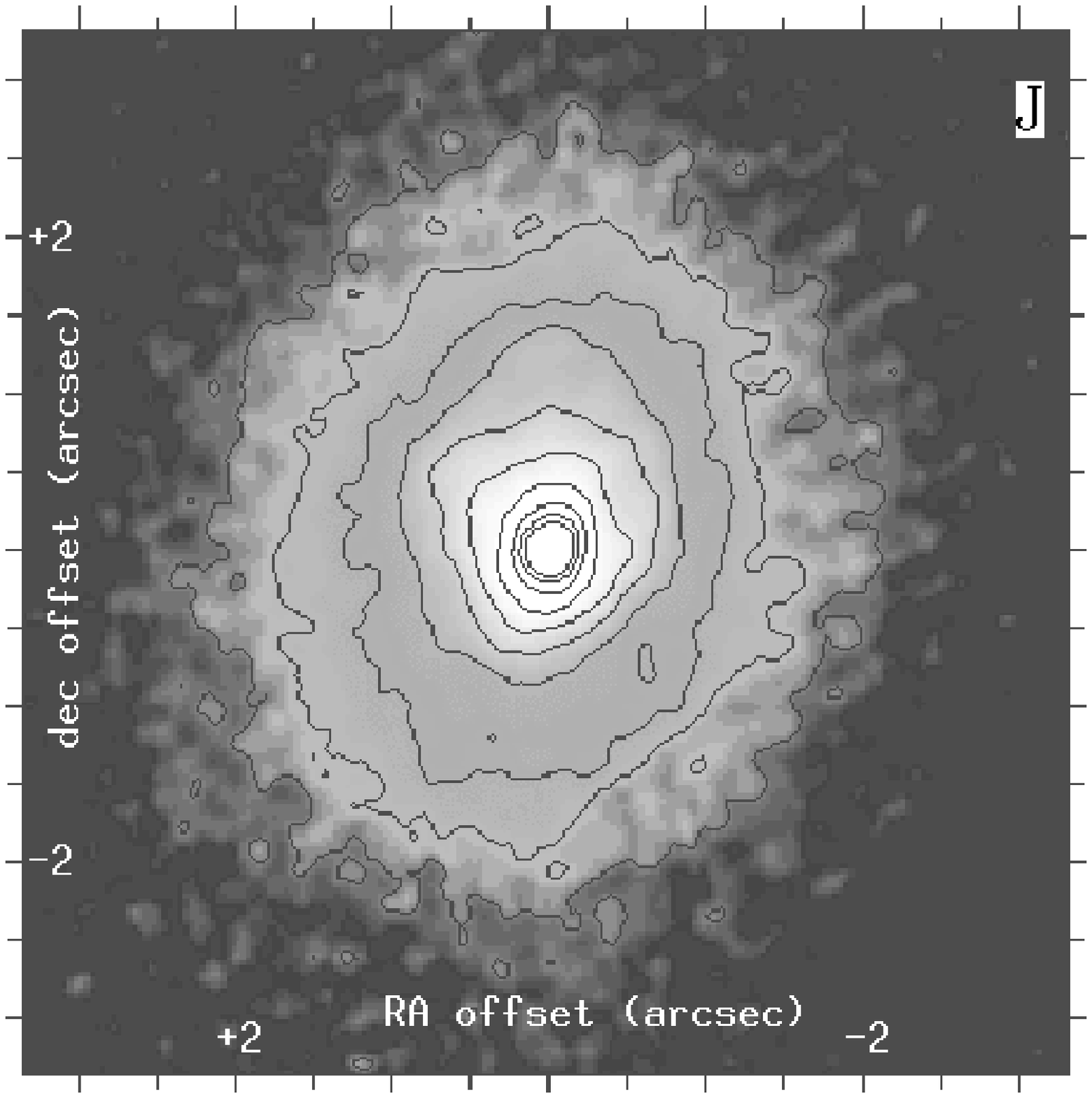, width=8.25cm, angle=0}
\end{center}
\caption{The central 6.7\arcsec\ of NGC\,3227 in $K$, $H$, $J$, F606W (hereafter $V$-band).
In all images, North is up, East is left.
The strong SW/NE feature in the HST $V$ image is 
due to bleeding/saturation in the nucleus.
} 
\label{fdeconv}
\end{minipage}   
\end{figure*}

%
% FIGURE 2
%
\begin{figure*}
\begin{minipage}{170mm}
\begin{center}
\epsfig{file=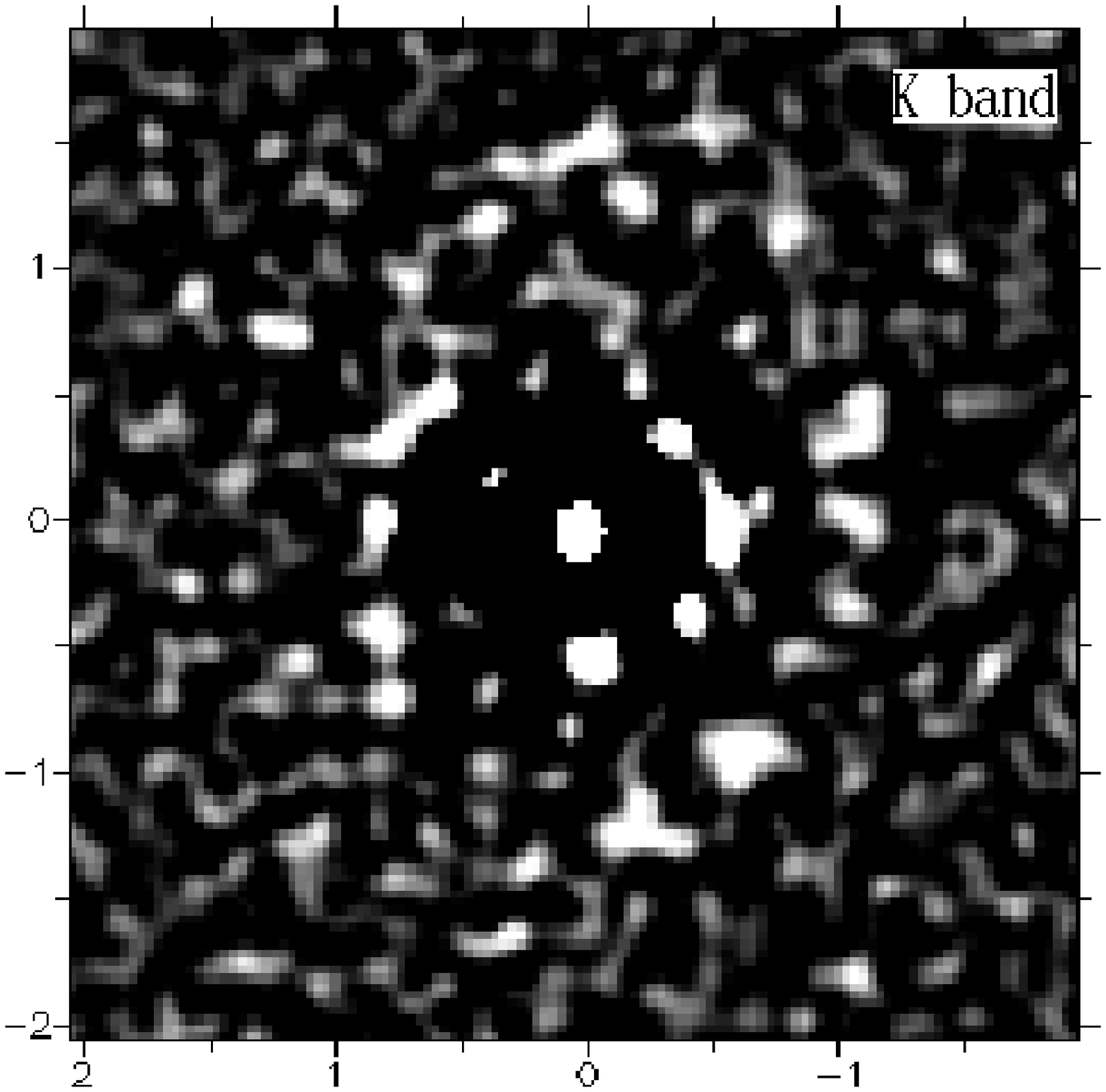, width=7.25cm, angle=0}
\epsfig{file=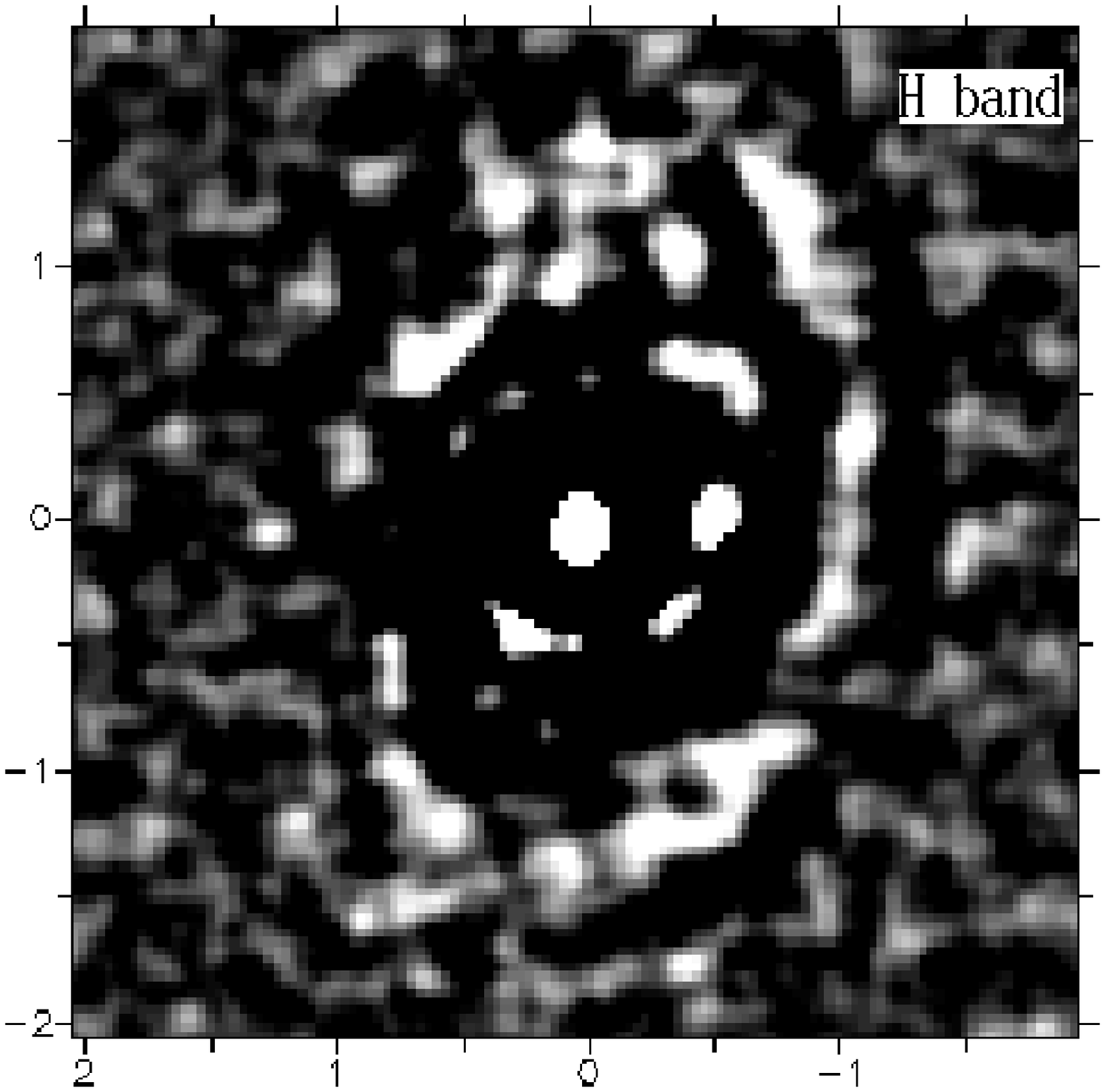, width=7.25cm, angle=0}
\epsfig{file=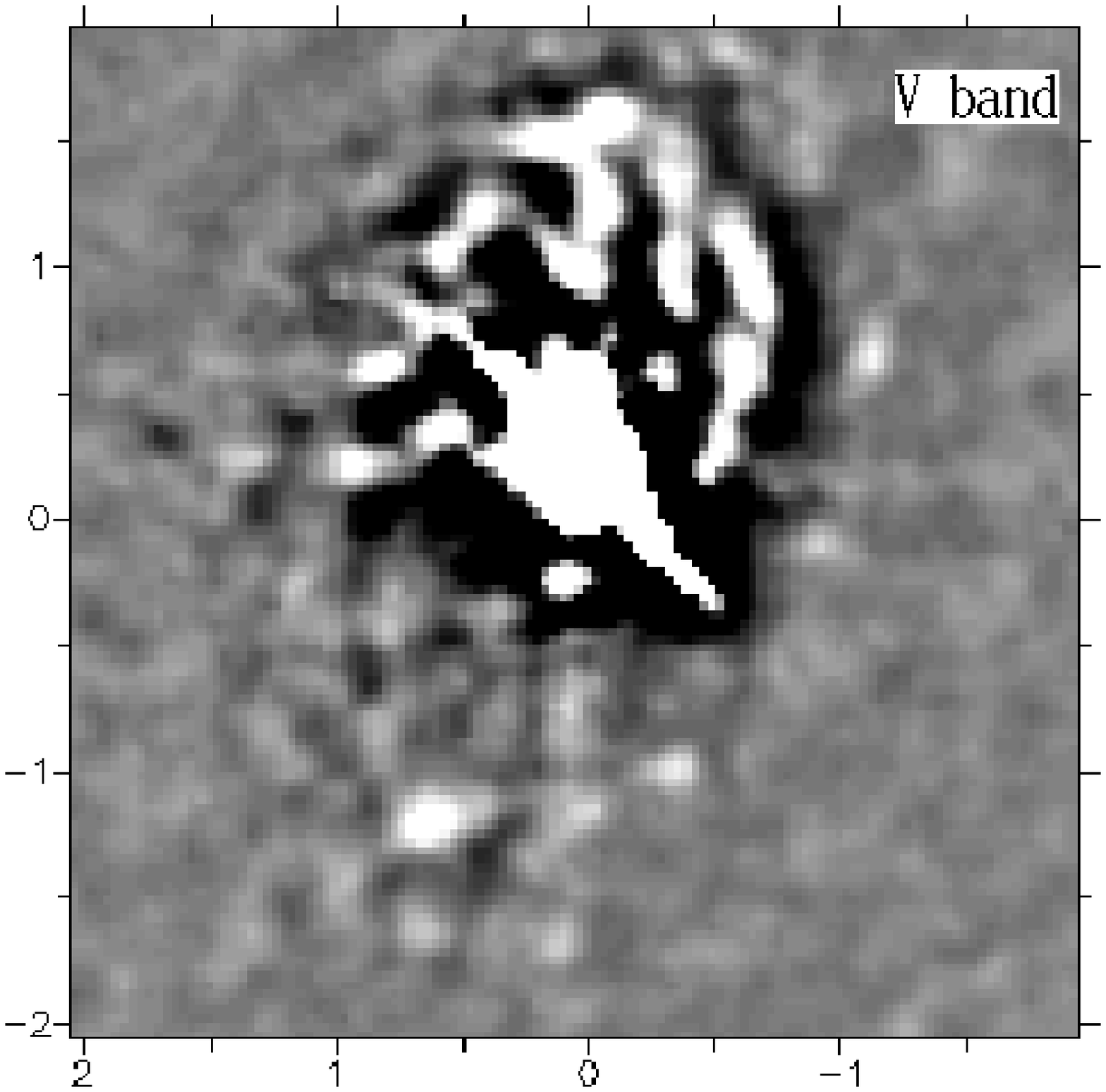, width=7.25cm, angle=0}
\epsfig{file=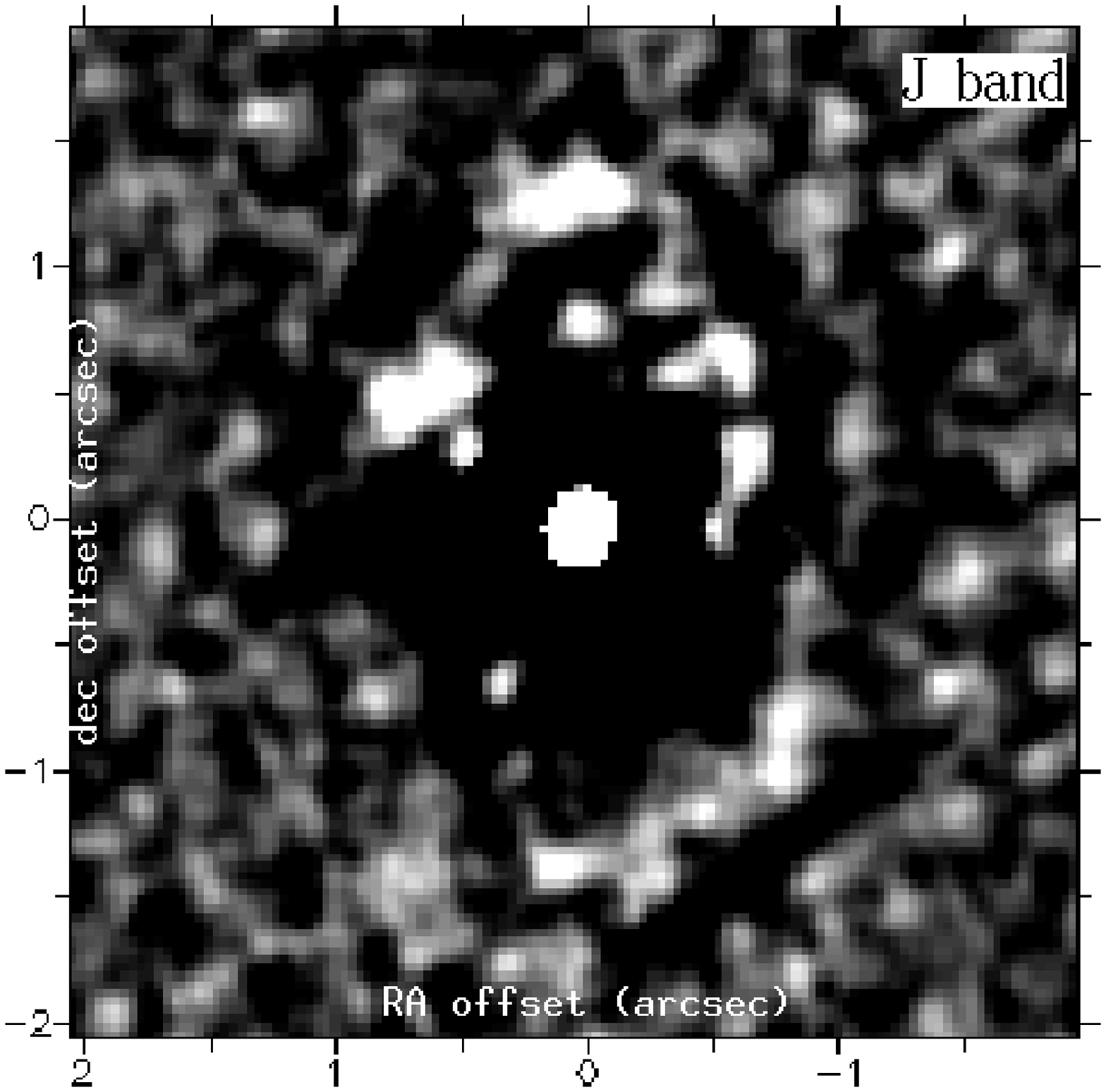, width=7.25cm, angle=0}
\end{center}
\caption{Close-up of central 4\arcsec\ region with smooth model subtracted 
images of NGC\,3227. Clockwise from upper left is $K$, $H$, $J$, $V$.
See figure 8 for a cartoon of the claimed structure.
The strong SW/NE feature in the HST $V$ image is 
due to bleeding/saturation in the nucleus.
}
\label{fsubtract}
\end{minipage}
\end{figure*}

%
% FIGURE 3
%
\begin{figure*}
\begin{minipage}{170mm}
\begin{center}
\end{center}
\caption{Colour maps of NGC\,3227, {\bf top row:} $J$-$K$, $H$-$K$,
{\bf middle row:} $V$-$K$, $V$-$H$, {\bf bottom row:}  
$V$-$J$. The colour scheme used has redder colours darker, and bluer colours
lighter.
The strong SW/NE feature in images formed with the $V$-band is 
due to bleeding/saturation in the nucleus of the HST $V$-band CCD image.
A blue ring at $\sim$2\arcsec\ radius is apparent in the $J$-$K$, $H$-$K$
maps, however the effect is largely obscured with the dusty optical HST image.
}
\label{fcolmap}
\end{minipage}
\end{figure*}

Colour maps are formed by smoothing the images to the worse resolution of
a given pair and taking the flux ratio (figure 3). %\ref{fcolmap}). 
The colour scheme used has redder colours darker (red) to emphasize dusty structures.
Any colour gradients in these images  can result from several different
processes: 
1) change in the optical depth from dust 
2) change in the stellar population 
3) change in the gas emission features (primarily \ha).
The most prominent feature is an irregular-shaped patch to the southwest.  
The fact that this region appears clearly as a deficit in the $V$-band image, 
and takes on a
patchy morphology is strong evidence for dust obscuration as the source of the 
colour gradient. The  region is therefore  most pronounced in the $V$-$K$ 
colour map, since the
$K$ image is least affected by dust. The $J$-$K$ and $H$-$K$ images
indicate that substantial dust still affects this region in the $J$, and even 
$H$-bands. 
We estimate A($V$)$\sim$8 magnitudes in this region, assuming an intrinsic
stellar index of H-K=0.22, J-H=0.78 (Glass \& Moorwood 1985), and an extinction
law A($K$)=0.112\,A($V$), A($H$)=0.175\,A($V$) and A($J$)=0.282\,A($V$) 
(Rieke \& Lebofsky 1985).
The colour maps reveal that the nucleus itself is very red, consistent with a reddened nuclear continuum component, with possible
thermal dust emission in the $K$-band.
The red colours surrounding the region of bright pointlike objects described 
above %of the  knotty spiral starburst 
stands out from a region
slightly bluer than the larger scale bulge of the galaxy.
To the northeast, the colours are generally bluer, although 
this may simply be in
contrast to the very red colours of the SW dust absorption. 
The $J$-$K$, $H$-$K$ maps reveal that there may be a blue ring with a 2\arcsec\
radius.

Although the HST $V$-band image has strong CCD bleeding from the bright nucleus,
a northeastern nuclear extension of about 1\arcsec\ to the isophotes is 
apparent in all the images except $K$-band (figure \ref{fdeconv}). 
It extends furthest, 
and with the largest position angle (PA), 
in the shorter wavelength $V$- and $J$-band images. 
The feature appears as an ellipse of $\sim 1$\arcsec\ extent 
in the colour maps, roughly orthogonal to the plane of the galaxy.
This $\sim$45$^\circ$ PA feature, has
bluer colours ($H$-$K$ = 0.10) than the surrounding regions, including the 
blue region ($H$-$K$ = 0.15)
surrounding the pointlike objects visible in the unsharp masked images 
(figure 2).
Given the difficulty in extinguishing K-band light when shorter wavelength
light is prominent, a natural explanation is young hot stars near the
galaxy core.
 
%The colour maps formed with the $K$-band image ($V$-$K$ and $J$-$K$ in figure 3),
%reveal a 
The images are distorted by PSF artifacts in the central 0.5\arcsec\,, 
plus the above mentioned CCD bleeding in the HST image.
However, the extent and position angle  (PA) of 43$^\circ$ indicate that this 
feature is likely 
unrelated to the AOB artifacts described in Chapman et al. (1999a,1998). 
Although the HST CCD bleeding lies along a similar PA,
the feature appears in all the near-IR colour maps by themselves.
Figures \ref{cartoona} and \ref{cartoonb} 
depict all of these structures relative to 
each other and the larger scale galaxy).

\subsection{Colour-colour analysis}

\begin{figure*}
%\begin{minipage}{170mm}
\begin{center}
\psfig{file=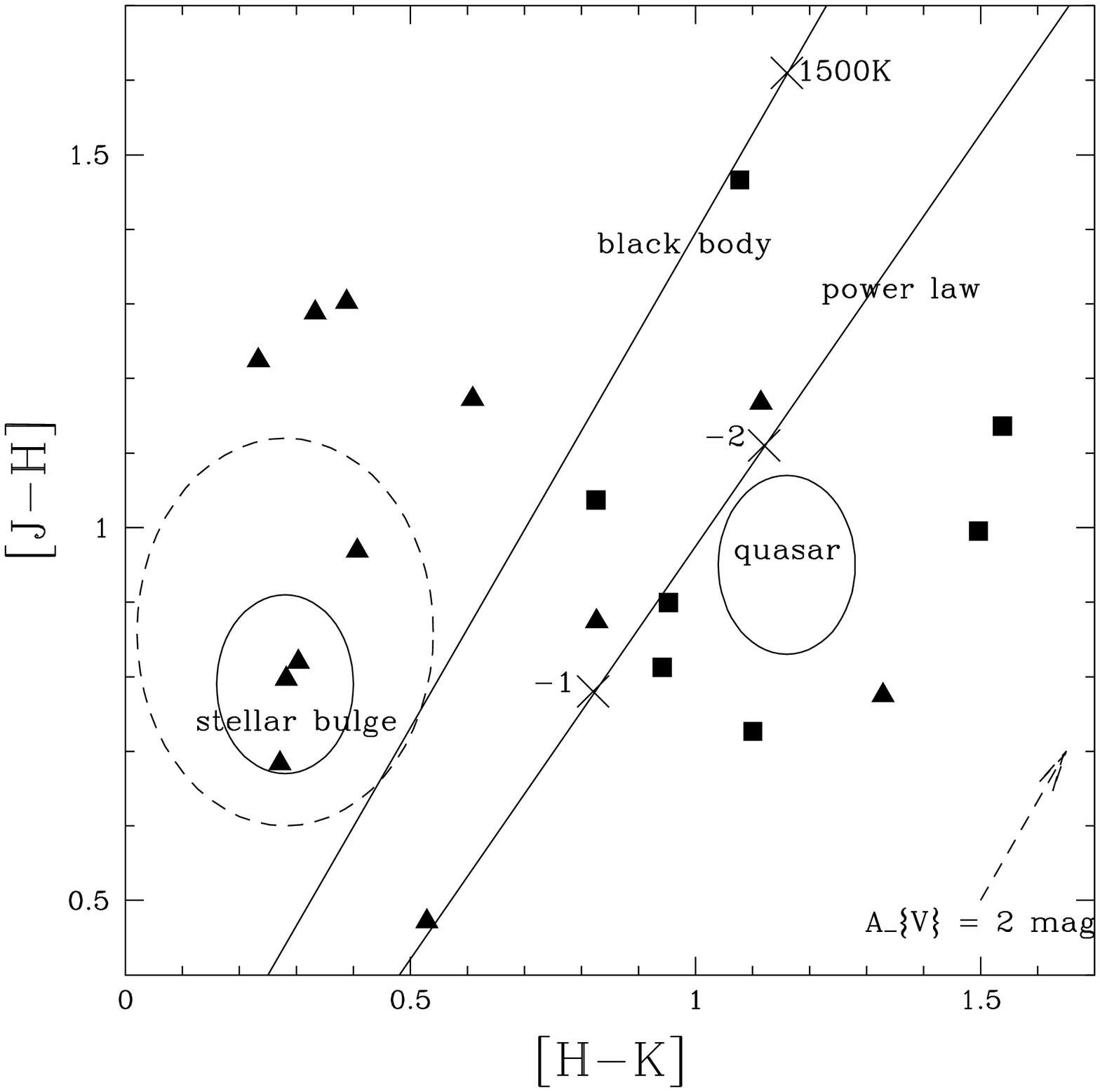, width=7.75cm, angle=0}
\end{center}
\caption{Colour-colour diagram of the knot-like structures  
in the core region. A vector showing the effects of 2 magnitudes
of  visual extinction is shown, along with the locus of colours for normal 
stellar bulges and continuum power law from quasar-type nuclei. The dashed
ellipse surrounding the normal bulge colours subsumes our error estimates
coupled with younger massive star colours.
The knots further from the nucleus ($>$0.5\arcsec\,) tend to have
colours near the locus of a stellar bulge (triangle symbols),
while those within 0.5\arcsec\ are more centered on the continuum power law 
region (square symbols).
%{\bf Right Panel}: 
%Chromatogram - a mapping of the colour-colour plane onto the image
%proper. Contours show NGC3227 $J$-band deconvolved image.
%Yellow represents normal stellar disk colours, while purple represents a
%supergiant population. The red colour of the nucleus is likely as a result
%of power law continuum plus dust reddening.}
}
\label{fcolourcolour}
%\end{minipage}
\end{figure*}

The colour-colour diagram of the point-like structures (hereafter called
knots) revealed in figure 2
is displayed in figure \ref{fcolourcolour}, highlighting the %18 
most clearly identified knots between model-subtracted $J$, $H$, and $K$ bands 
images (figure 2). The colours are measured through aperture photometry,
where apertures of 0.2\arcsec\,, 0.3\arcsec\ and 0.4\arcsec\ were used to 
derive an estimate for the uncertainty of 0.5 magnitudes for $J$-$H$ 
and 0.3 magnitudes for $H$-$K$.
The larger uncertainty in  the $J$-$H$ index is likely a result of the poorer
resolution in $J$-band relative to $H$ and $K$,
and some of the knot regions blend with surrounding ones.
The colours of the regions and visible knots were verified in the actual
colour maps, where the estimated photometric accuracy is $<0.3$ magnitudes.
Within the uncertainty there is still a rather large spread in the 
colours for these knot structures.
% as might be expected for the core region of a strong active galaxy. 
The knots further from the nucleus tend to have 
colours near the locus of normal spiral bulge stars (triangle symbols), 
while those closer in are more centered on the continuum power law region
(square symbols). This can be explained if 
scattered nuclear light and thermal dust emission are superposed with a 
starbursting component for the nuclear knots.

\subsection{Isophotal analysis and small-scale bars}

%
% FIGURE ISOPHOTE E,PA,SB
%
\begin{figure*}
\begin{minipage}{170mm}
%(a) \hskip 62.5mm (b)
\begin{center}
\vskip -5mm
\epsfig{file=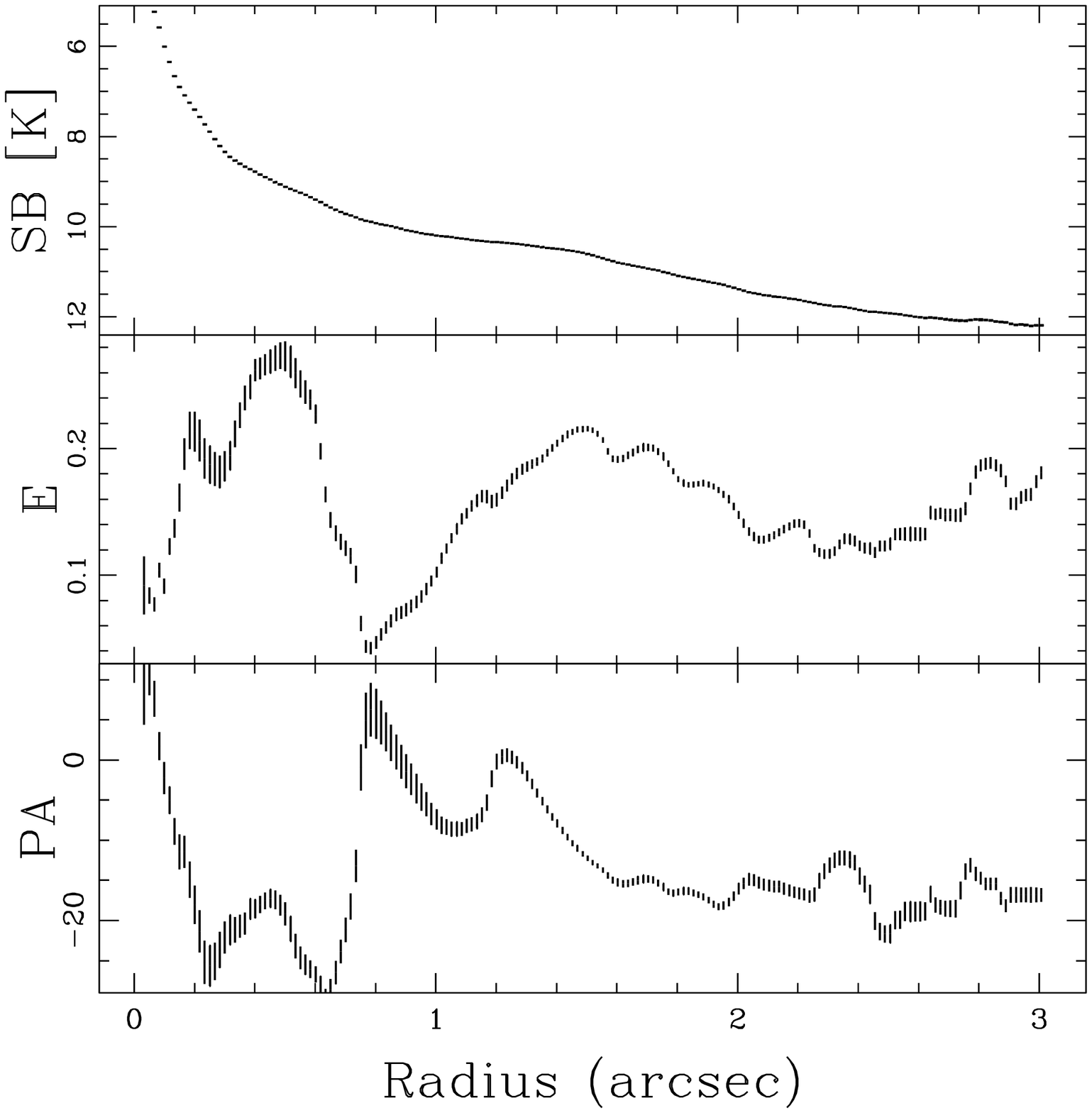, width=7.25cm, angle=0}
\epsfig{file=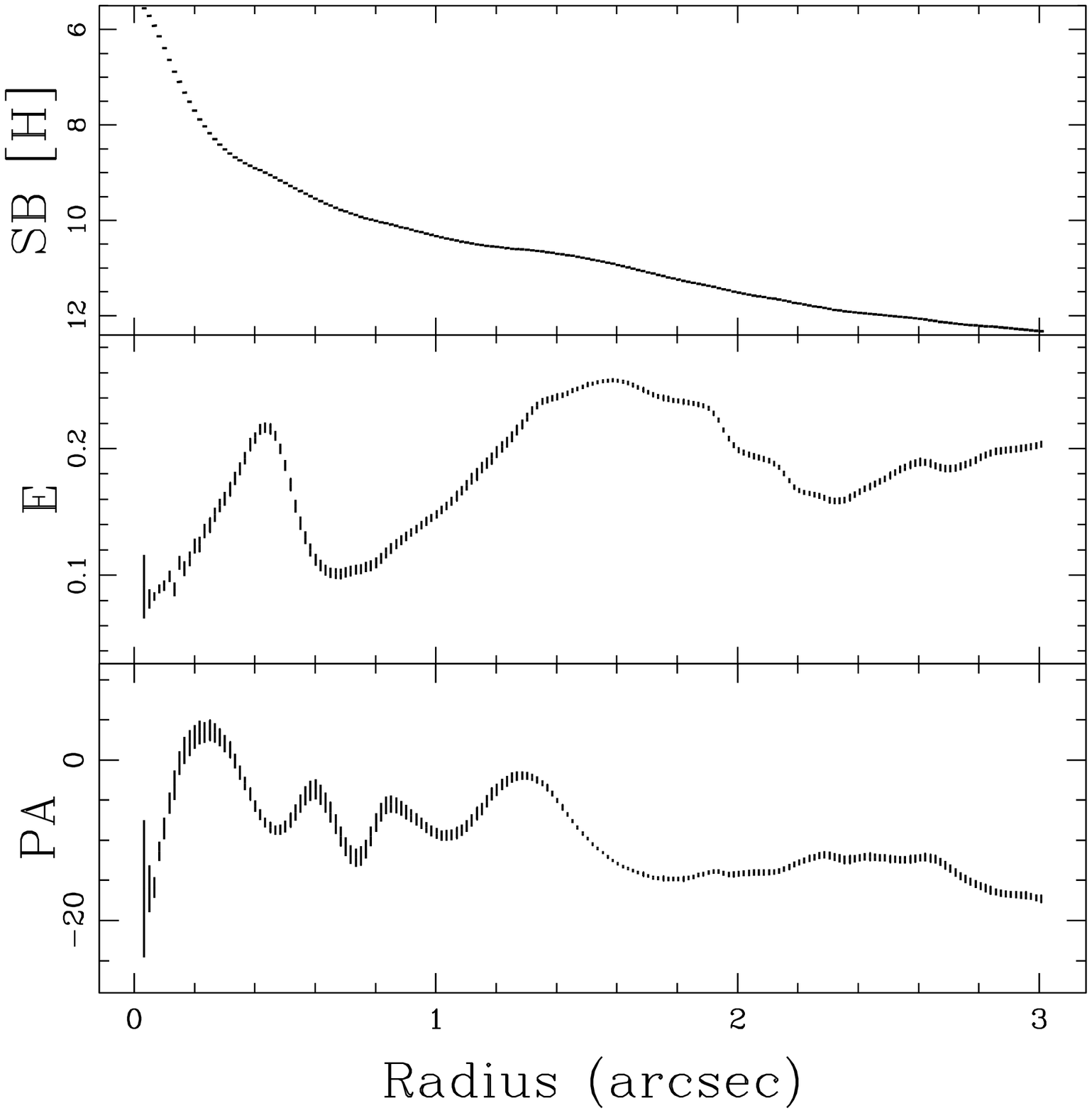, width=7.25cm, angle=0}
\epsfig{file=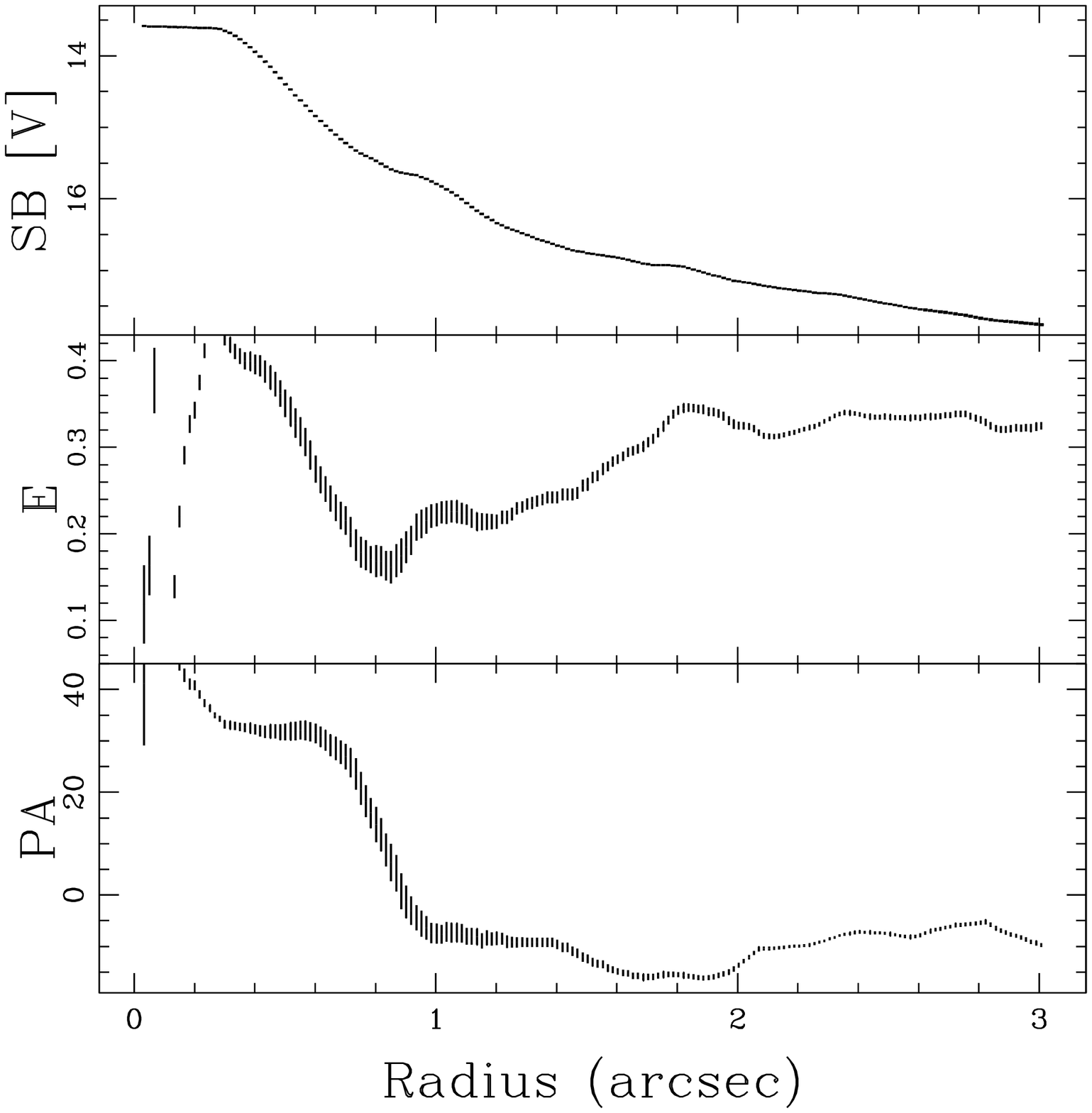, width=7.25cm, angle=0}
\epsfig{file=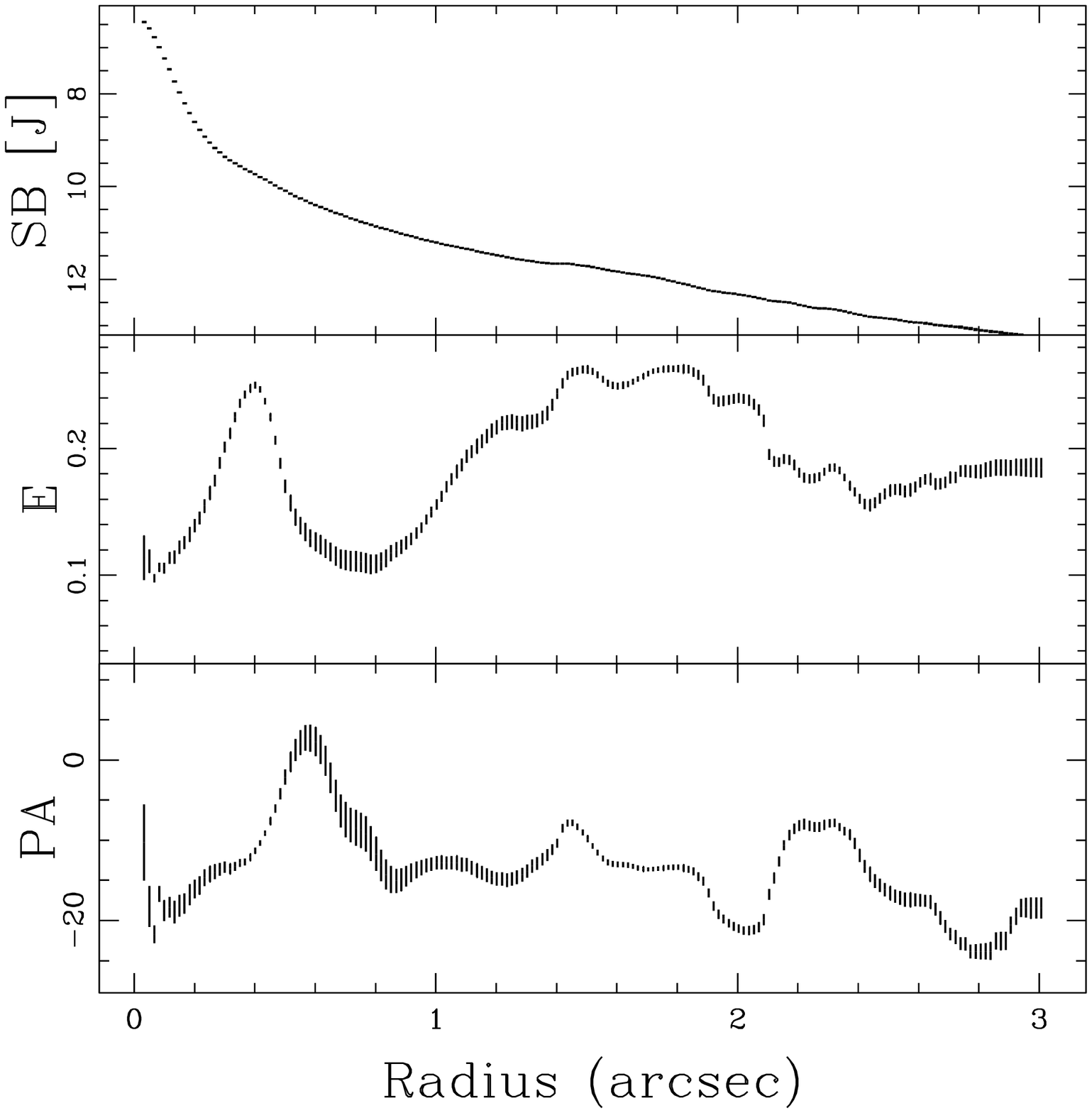, width=7.25cm, angle=0}
\end{center}
\caption{Profiles of NGC\,3227 from elliptical isophote fits. Clockwise from
upper left:  $K$, $H$, $J$, $V$. The panels (top to bottom) depict
surface brightness (SB) in mag/arsec$^2$, ellipticity (E) as $1-b/a$, 
and position angle (PA) in degrees from north. 
Note that the saturated HST $V$-band surface brightness is flat in the core. Error bars represent 1$\sigma$ isophote fitting error.
The varying position angle over the peaks in ellipticity suggests that there
is no clear small-scale bar potential in this Seyfert galaxy.}
\label{fprofiles}
\end{minipage}
\end{figure*}

The transfer of mass from large to small scales is a vexing problem in AGN
research.
From a theoretical standpoint, galactic bars are perhaps the most viable
candidate for facilitating this mass transfer (Schwartz 1981; Norman 1987;
Shlosman, Frank \& Begelman 1989).
The relatively constant surface brightness along the bar results in a 
clear signature in the radial profile analysis of a galaxy.
We follow the criterion for a nuclear bar discussed in
Friedli et al. (1996). This involves a rise then fall in ellipticity (E)  
while the position angle (PA) stays ~constant over the rising E. Double bars would appear
as two bumps in E, and a bar plus twist has
the PA and E (or just PA) changing simultaneously after the
bump. In the larger scale galaxy, the ellipticity would 
then decrease to
reveal the inclination of the disk, but our field does not extend this far.
If there is more than a 10 degree shift in position angle over a bump in E, 
then the feature is classified as  a twisted isophote rather than a bar.

The variations of surface brightness (SB), ellipticity (E), and position angle
(PA) for the galaxy were extracted from the fitting of elliptical 
isophotes. The profiles are similar at $J$, $H$ and $K$ 
(figure \ref{fprofiles}),  
displaying bumps in ellipticity at  0.5\arcsec\ and 1.5\arcsec\ radius, 
confirming the 
presence and position angle of the nuclear ellipse  (PA $\sim40^\circ$) 
and the larger enhanced region coincident with the spiral
starburst (PA $\sim -10^\circ$). 
The  $H$ and $K$ bands are less obscured by dust
and the profile is most clearly defined at these wavelengths, whereas the 
large amount of extinction to the southwest in the $V$-band skews the
isophote E and PA fits.  

For the spiral knot region ($\sim$1.5\arcsec\ radius), 
the isophotes are twisted of order 10 degrees. 
The images (figure \ref{fdeconv})  
show that the region extends to the south of the 
nucleus at $H$ and $K$ bands 
and would be consistent with a bar potential by our 
above criterion. At $V$-band, the region only appears to extend to the
north due to the southern dust obscuration.
 
For the nuclear ellipse, the PA twists more  significantly ($\sim 20^\circ$), 
and its twist and orientation vary with wavelength. At the smallest
scales, this is likely the result of  the wavelength dependent AO/PSF artifacts
dominating the signal. For the outer extent of the disk, the multiple emission components possibly
contributing to the flux (see  above) likely vary with wavelength, with
a strong stellar component at a larger PA in the $J$-band.
The twisted isophotes do not satisfy our bar criterion,
but the profiles are difficult to analyze with 
certainty as the feature is near the limit of our image resolution and clouded 
by artifacts in the very core regions. 

%Higher resolution images may reveal an even smaller-scale nested bar within this region.

\subsection{Comparison with MERLIN radio data}

%
% FIGURE RADIO
%
\begin{figure*}
\begin{center}
\vskip -5mm
\epsfig{file=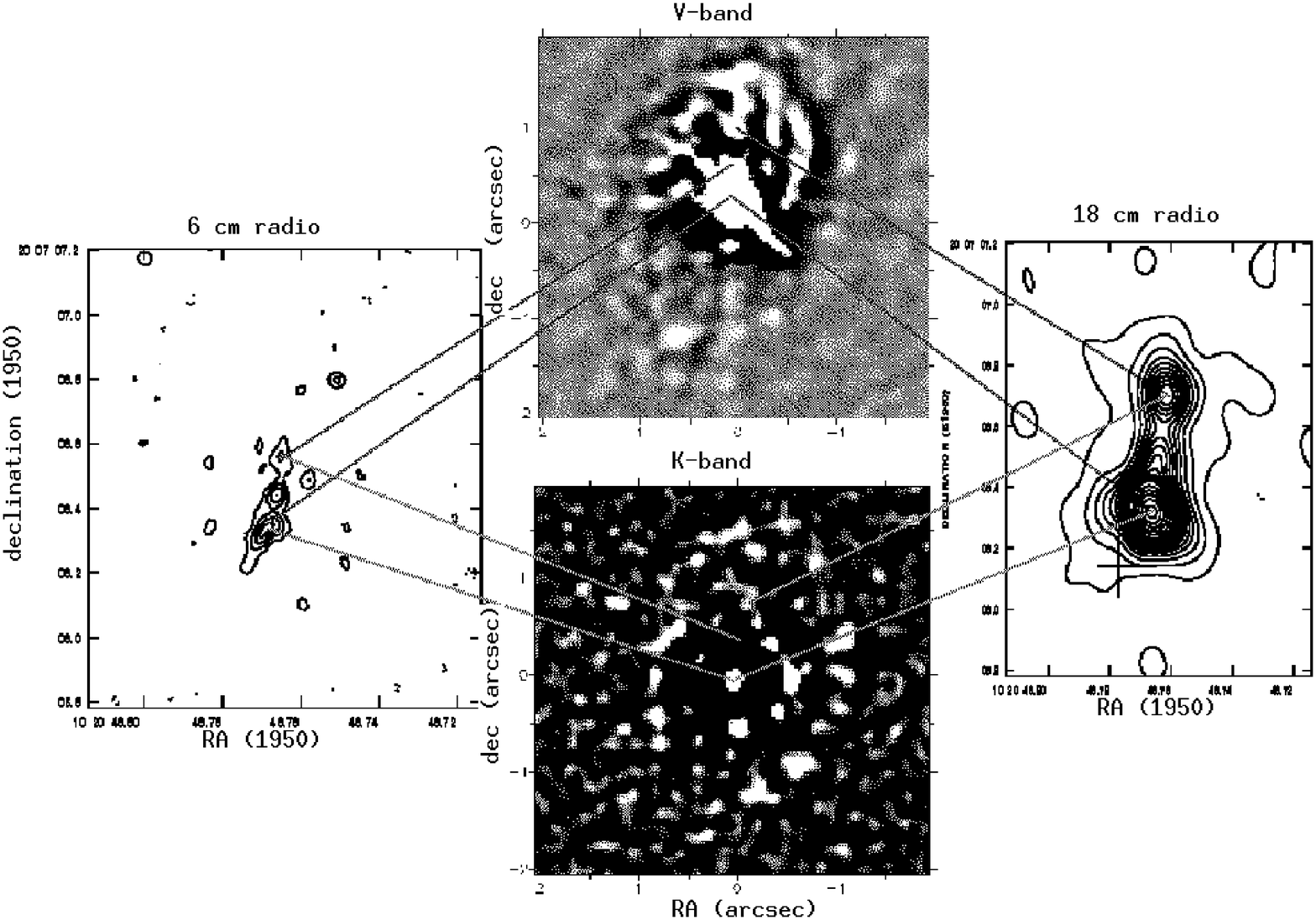, width=16.25cm, angle=0}
\end{center}
\caption{The 6\,cm (left contours) and 18\,cm (right contours) of NGC~3227 showing
the location in the model subtracted V (above) and K (below) bands. Note
that there is close alignment of the peaks of the radio maps, with the galaxy
core and knot to the north.}
\label{fradio}
\end{figure*}

The images can be compared to the 6\,cm and 18\,cm MERLIN radio continuum 
emission (Courtesy of C. Mundell. Originally published in 
Mundell et al. 1995), which 
align with the axis of the nuclear spiral (figure \ref{fradio}). 
Previous explanations for the
radio structure (Mundell et al 1995, 1996), 
invoked the standard unified AGN model
to explain this emission. In this standard unified model, a dusty
torus preferentially blocks certain views of the active nucleus, and
results in anisotropic emission of ionizing continuum photons. The result
is a cone-like region of ionized gas, such as [OIII]. 
Even at the low resolution of their optical images,
it was apparent that there was an offset in orientation from an [OIII] ``cone"
and the small-scale radio features. 
A projection effect would be possible, but this would necessitate that the
NE side of the disc is closer to us than the SW side. This would only be
possible if the spiral arms were leading rather than trailing (Mundell et al. 1995, Pedlar et al. 1987).

Our high resolution imaging suggests that 
the radio emission may be unrelated to any kind of jet outflow, and instead is
associated with the nuclear knots observed in figure 2. 
The peak to the north is only 
observed at 18cm with no 6cm counterpart, possibly indicating a steep spectrum
consistent with supernova remnants in the regions of star formation. The 
southern peak spectral index is consistent with a compact active nucleus 
(Mundell 1995, Ho 1999). The absolute positional accuracy of our AOB images
was limited to the pointing accuracy of CFHT, and is obviously 
worse than the VLA.
An error of a few tenths of an arcsec is possible in the relative image
alignments. 
We align our near-IR nuclei to the brightest radio peak. Figure \ref{fradio} 
indicates that the radio then aligns well with both the orientation and some
of the knots of the spiral/ring assembly, and a possible interpretation is that
we are seeing synchrotron emission from SNe remnants.

%Mulchaey N2110: Jostling VLA contours is fair.

\subsection{Unified models and OASIS spectroscopy}

The OASIS instrument on the CFHT allows spatially resolved spectroscopy to be obtained
at the resolution provided by the Adaptive Optics Bonnette.
Spectral imagery was obtained in the \ha\ and [SIII] lines under fairly
poor seeing conditions and only part of the data obtained is usable since
guiding was lost intermitantly in some of the exposures. The best corrected FWHM
obtained is only 0.8\arcsec\ using the
0.3\arcsec\ per pixel sampling.
The [SIII] lines are excited under similar conditions to the [OIII] line and
is typically 1/3 the peak intensity. The advantage for adaptive optics is
that the PSF correction is much better at 9000\,\AA (SIII)  than 5000\,\AA (OIII).

Initial results from our OASIS data (Chapman \& Morris, in preparation) seem to be consistent with the
unified model for NGC\,3227 proposed in \cite{gonz97,ar94}
providing a higher resolution perspective on the HII region to the
southwest, as well as the extended [OIII] and \ha\ to the northeast.
The dusty region in the $V$-$K$ color map falls over the blank region which separates the HII region from the
core (Figure \ref{cartoonb}), likely explaining the disconnected morphology.
In the unified AGN picture, biconical emission line regions are thought to be the
result of collimated continuum emission via a nuclear dust torus. Although with NGC\,3227 one would expect a counterpart to the
extended [OIII] emission ``cone'' on the opposite side of the nucleus (SW),  the
 dusty region would
obscure such emission in addition to any \ha\ that extended from the nucleus
to the separated HII region.
%There is however some evidence for smaller-scale conical morphology in our [NII
% / \ha\ ratio map, with a larger ratio along the NE--SW axis defined by the
% [OIII] emission (figure \ref{fratio}).

However, a larger scale extended [OIII] region (Arribas \& Mediavilla~1994, 
Mundell \etal 1995, Gonzalez-Delgado \& Perez~1997)
lies to the northeast and has been
interpreted as a narrow-line region ionized by the AGN, collimated into
a cone by a small-scale ($\sim$ pc) dusty torus.

%\begin{figure}
%\begin{center}
%\epsfig{file=figs4b/ha_rat.eps, width=8.25cm, angle=0}
%\end{center}
%\caption{
%North is up, east is to the left}
%\label{fratio}
%\end{figure}

%****************** MAIN DISCUSSION **********************
\section{Discussion}

\subsection{Fueling the nucleus}
%A succession of bars? No Big bar then mini spiral

We can then trace the structural components of this galaxy from the
largest scales down to the small scales observed in our images
(listed in Table 1; cartoons in figures \ref{cartoona} and
\ref{cartoonb} show a depiction of scales and relationships of the 
various features).
%Several possible scenarios emerge from these results.
On the largest scales
Gonzalez-Delgado \& Perez~(1997) 
noted that a large-scale bar appears to transport
material towards an inner radius which corresponds to the calculated Inner
Linblad Resonance (ILR) at roughly 7\arcsec\,.
At this point, prominent dust and HII regions indicate
substantial star formation.
Within this region, non-circular motions are observed in the CO rotational 
transitions by Schinnerer \etal (1999, 2000),
using millimeter interferometry at IRAM.
The gas motions can be successfully described by a molecular bar or
%of length $\sim1$kpc 
by circular motion in a tilted ring system, characteristic of a 
warped disk (Schinnerer \etal 2000). 
The apparent alignment of a possible CO bar with the larger optical
bar may represent a coupled structure, rather
than two kinematically distinct bar potentials (see Friedli 1996,
Friedli \etal 1996, 1993 for discussion of nested bar alignment).

The calculated ILR of $\sim$2\arcsec\ for a molecular bar model
corresponds to the outer extent
of our observed ring/spiral of knots (row (7) in Table 1 -- see 
model subtracted images, figure 2)
and the blue ring visible in the near-IR colour maps (row (6) in Table 1 -- 
see figure 3).
Whether the non-circular molecular gas motions are due to a warped disk or bar,
the collected gas at these radii (2\arcsec\,) 
implies recent star formation. 
The colours of our outer
ring/spiral of knots,
and the blue $J$-$K$ ring (2\arcsec\ radius) 
are consistent with such a star forming scenario.

Our observed smaller scale blue annulus ($J$-$K$ map --
row (10) in Table 1 -- see figure 3) 
extending to a radius of 0\farcs5
lies within the warped CO disk in Schinnerer et al.~(1999).
Our $J$-$K$ and $H$-$K$ maps (figure 3) 
show a red peak 1\arcsec\ east of the nucleus
which matches the CO intensity peak in Schinnerer et al.~(1999), consistent
with reddening from the molecular gas. 
The molecular gas in this region is observed to be counter rotating
(Schinnerer et al.~2000), consistent with the warped disk becoming
perpendicular to the plane of the galaxy, and
our small scale blue annulus may form an inner region of the warped CO disk.

With the bar structure existing at larger scales and extending down to
the knotty spiral at 2\arcsec\ in our images, along with 
the torques induced by an
even smaller scale warped disk, it is natural to speculate that 
the small scale features observed in our images
may provide an indication of how material is funneled down to the scales
($\sim$ pc) where viscous forces may take over to fuel the AGN.
With a profile analysis (figure 5), 
both the region of spiral knots and the smaller
scale elongation seen in the colour maps are identified as ellipticity peaks.
The large amount of twisting in position angle over these peaks indicates
that the features are likely not associated with nested bar-potentials. 
%ACTUAL FORMATION OF NUCLEAR SPIRAL ARMS?
In the absence of any true bar potential in the nuclear region, the fate of
knots of star formation generated near the 2\arcsec\ ILR would then be to
slowly drift inwards with
time (Morris \& Serabyn 1996), being carried apart by differential rotation.
This process could mimic the spiral/ring assembly %association?
seen in the core region.
%NOTE projection could make it appear with any ellipticity.

The extended radio feature supports such a fueling picture.
Dense blobs spiraling in from 100's of parsecs due to a viscosity provided 
by a hot outflowing may provide a means to increase the inflow velocity.
However cloud stability is clearly a major problem in this case.
In our own galaxy at roughly
comparable scales, similar processes are thought to be at work (Morris \etal 1996).
Although the Milky Way is thought to have a supermassive black hole in the core
(Ghez \etal 1998),
it appears not to be active currently, possibly due to lack of fuel, or
a means to shorten the inflow timescale.

%and the
%explanation for why some galaxy cores are fueled and some
%not would have to lie at even smaller scales.
%The possibility also
%exists that the activity is simply at a low level
%sag A?

In certain Seyfert galaxies, Regan and Mulchaey (1999) 
have found small-scale spiral dust
lanes that may be able to provide fuel for the central engine.
This is consistent with the results of Chapman et al. (1999c).
This suggests that material may indeed
be transported inwards by the spiral potential in many AGN, 
rather than losing its angular momentum
in the presence of a strong bar as hypothesized by Shlosman et al. (1989).
Our images of NGC\,3227 with the spiral knot region
would be consistent with this picture.
% QUES: seen as a red spiral in the colour maps (figures 3). %\ref{fcolmap}).
Although it is clear that spiral potentials can transport material 
inwards (Binney and Tremaine 1987) 
there has been little theoretical or numerical work to date
to support such
a scenario. It is not yet clear that the timescales involved could
effectively fuel the active nucleus.

A contrary point concerns
the fact that the alignment with the
axis of the feature described as an [OIII] ionization
cone (Gonzalez-Delgado et al. 1997)
indicates that our small-scale
blue annulus (Table 1, row 10) may represent scattered AGN light.
%This is made all
%the more convincing by the blue colour of our northeast elongated region.
On the other hand, if our observed small-scale elongation is a
twisted disk or torus, similar to that found in Centaurus\,A by Schreier et
al.~(1998), its plane lies roughly perpendicular
to the axis defined by the radio ``jet" observed at 6 and 18\,cm, and would
be consistent with a collimated radio jet normal to an accretion disk plane.
For the radio emission to be interpreted as an outflow,
the collimated ionization of the NLR picture
would then likely have to be abandoned.
%NOTE THAT GDM IS FINE WITH THIS PICTURE
We explore these scenarios below.

%
% TABLE STRUCTURES
%
\begin{table*}
{\scriptsize
\begin{center}
\centerline{Table 1}
\vspace{0.1cm}
\centerline{The structural components of NGC\,3227, from largest (1) down
to the smallest (10)}
\vspace{0.1cm}
\begin{tabular}{llllll}
\hline\hline
\noalign{\smallskip}
{Component} & {Scale} & {PA ($^\circ$)} & {E (1-$\frac{b}{a}$)} &
{Observed with} & {Function in galaxy} \cr
\hline
\noalign{\smallskip}
(1) Large Scale galaxy & 1-10 kpc & -22 & 0.3 & V-band & \cr
(2) Large Scale bar & 1-5 kpc & -20 & 0.5& galaxy subtracted & 
	funnel material to ILR at 7\arcsec\ \cr
(3) Extended [OIII] & 1kpc & 35 &  & OIII filter/ OASIS [SIII] & collimated emission
? \cr
(4) Circum-nuclear ring & 1 kpc &  &  & \ha\ & ILR\cr
(5) Medium scale bar & 100-1000pc & -16 &  & submm CO  & funnel material to ILR at 2\arcsec\ \cr
\noalign{\smallskip}
(6) Blue ring & 300 pc & n/a & 0 & colour maps, esp.~$J$-$K$ &
young stars? \cr
(7) knots in spiral/ring & 200 pc & -10 & 0.2 & model subtract, colour maps&
spiral starburst, ILR\cr
(8) $V$-$K$ ellipse & 200 pc & -10 & 0.1 & $V$-$K$ map, raw $V$ & bluer than galaxy \cr
(9) Radio jets/blob & 100 pc & -10 &  & MERLIN 8/16cm& SN/outflow?\cr
(10) $J$-$K$ annulus & 100 pc & 40 & 0.3 & all colour maps \& images & twisted disk/bars, scattered AGN light \cr
\noalign{\smallskip}
\noalign{\hrule}
\noalign{\smallskip}
\end{tabular}
\end{center}
}
\end{table*}

%
%%% FIGURE CARTOON A %%%
%
\begin{figure*}
\begin{center}
\end{center}
\caption{Palomar Digital Sky Survey (DSS) image of NGC\,3227 interacting with the dwarf elliptical
companion NGC\,3226 to the North. 
Inset cartoon of the larger scale features: the large
scale stellar bar funnels material to the ILR (depicted by the large circular
region). A proposed smaller-scale molecular bar or warped disk
(Schinnerer et al.~2000) resides within (PA is
approximate), presumably transporting
material down to the scales probed with our AO images (overlapping ellipses in
the very core). 
The [OIII] ionization cone and detached HII region are shown to scale. }
\label{cartoona}
\end{figure*}

%%% FIGURE CARTOONB %%%%
\begin{figure}
\begin{center}
\psfig{file=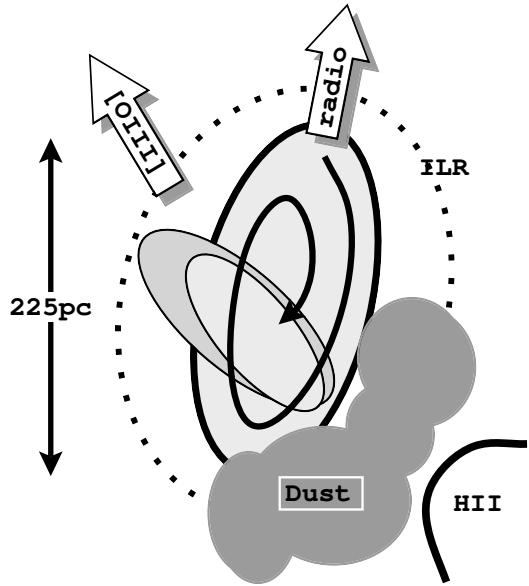, width=7.25cm}
\end{center}
\caption{Cartoon of the smaller scale features probed by the CFHT AOB. 
The ILR of a possible molecular bar
(Schinnerer et al.~2000) 
is depicted as the dashed ellipse surrounding the spiral
starburst region. The radio ``jet" (PA $= -10$) aligns with this spiral region.
The dusty region to the SW may be a reason for
the lack of [OIII] and the detached HII region. The two possible components
of the inner nuclear disk/torus 
are shown as overlapping ellipses (larger PA of
the $J$ and $V$-band isophotes, and smaller PA of the $H$,$K$-band isophotes). 
This may correspond to the inner region of the warped molecular disk
in Schinnerer et al.~(1999). 
The direction 
of the extended [OIII] region is shown by an arrow.
The solid black (one-arm) spiral is drawn to suggest the possible structure
associated with the nuclear knots seen in figure 2.}
\label{cartoonb}
\end{figure}

\subsection{Unified models and the role of the radio feature}
\subsubsection{AGN origin to the radio feature}

%NOTE just say the resolution and colors are not sufficient to
%distinguish the role of the extended isophotes = nuclear disk/torus etc.

The radio extension may be a jet powered by the Seyfert nucleus. 
We could then interpret the twisting
nuclear ellipse in the central 2\arcsec\ 
as a warped disk/torus related to the collimation axis defined by the radio 
feature, which we refer to as the {\it jet} in this section. 
Several arguments support this picture: 
{\it (i)} there is a 
molecular disk counterpart %(although almost unresolved at 1\arcsec\,) 
at a similar scale with apparent counter rotating gas
(Schinnerer et al.~2000);
{\it (ii)} the near-IR disk/torus is roughly perpendicular to the jet; 
{\it (iii)} the twisting of the disk/torus is in the orientation such that the
structure 
would be more perpendicular with the jet at the smallest scales.
The feature might even be associated with an extended torus as seen for 
instance in NGC\,4261 (Jaffe \etal 1996), and as described in the
models of Maiolino \etal (1997). 

The radio jet aligns well with the north-south elongated region of our
knots, and the radio peaks may correspond
to knot features. We could interpret this bluer  
region as a narrow line emission region (NLR) excited by the AGN 
continuum. Kotilainen \etal (1997) also found a blue excess in their 
optical colour maps lying in two lobes surrounding the nucleus on the NW-
SE axis which they interpreted as scattered nuclear light. Higher 
spatial resolution spectral imaging is required to confirm whether the
line ratios and dynamics within this region are consistent with
the continuum ionization of a NLR.
%different kinematic state

The suggestive spiral/ring morphology of the knots (figure 2) makes it 
less likely that this region is a random assembly of starburst knots 
resulting from shocked gas in the NLR/radio jet entrainment.
In addition, the nuclear disk may have two emission components at different
PA associated as suggested in section 3. Although, within the disk, the 
more extended ellipse at larger PA certainly appears stellar in 
origin, the inner disk at smaller PA may be substantially scattered 
AGN light, implying that the radio jet does not necessarily 
define a unique collimation axis. 

However a small scale ($\sim$ pc) dust torus which collimates the larger scale 
emission is not a mandatory feature of AGN. Indeed Malkan \etal (1998) have
put forth a viable alternative scenario where patchy dust up to 100\,pc from
the nucleus would lead to the classification differences seen in many AGN
(a Galactic Dust Model -- GDM).
Our near-IR extinction maps for a large sample of Seyfert galaxies show that 
this is a plausible picture (Chapman \etal 1999c).
Cloud-cloud interactions associated with the dense knots observed in our
central region may effectively provide 
a quasi-spherical cloud distribution in this region, providing a
natural setting for this model put forth by Malkan \etal 1998.
If the GDM describes NGC\,3227, then there is not necessarily a 
contradiction in radio/NLR misalignment: 
the radio jet extends along the presumed spin axis of the AGN,
%REFERENCE?
while the AGN continuum excited emission (NLR) extends anisotropically where it
is not obscured by patchy dust -- primarily towards the NE.

%NOTE Delgado bar has blobs along the spiral ellipse axis.
%This may be an effect of the large scale bar at small R?

\subsubsection{Starburst origin to the radio feature}

Regardless of whether the extended [OIII] narrow-line region to the northeast 
defines a collimation axis of a small-scale obscuring torus, there does
seem to be evidence for preferential continuum ionized emission towards
the NE regions.  The scattered light interpretation
of at least part of the disk feature is consistent. 

A case can therefore be put forth that there is a ``classical" bi-conical 
emission structure in this galaxy, largely obscured to the SW, and collimated
at small scales by some kind of dusty torus.
The elliptical region surrounding the starburst knots might then represent
a disk, orthogonal to the ionization axis, with well defined spiral arms 
and small-scale bar, facilitating mass transfer to the core region. 
With this geometry, in addition to the radio morphology extending along
the axis of the ellipse of knot structures (figure 6) (possibly with
knots coincident 
with the actual
radio peaks), the most plausible explanation
for the radio emission  seems to be supernova remnants associated with
the starforming clumps.  
If we accept this picture, it is difficult to interpret the radio feature as an
AGN-driven jet mis-aligned with the ionization cone, since 
projection effects are not consistent with the known motions of the 
galaxy (Mundell \etal 1995). 
It does not necessarily preclude an outflow driven by starburst 
superwinds (Heckman et al.~1990, Unger et al.~1987), since the superwind 
outflows are preferentially blown out of the plane of the galaxy, consistent
with the galactic motions with the SW side of the disk being closer to us than 
the NE side. 
An AGN collimation axis would not be expected to lie in any preferred direction
(Malkan et al.~1998) and there is no problem interpreting the ionized
regions within this geometric picture. 

%The  may still be a stellar bar/disk component (PA=40) perpendicular 
%to the galactic disk (PA=-20). 
%However, in this picture, the nuclear disk is likely not any kind of 
%extended torus as seen for instance in NGC4261 (Jaffe et al. 1996) 
%be in direct conflict with the ionization cone hypothesis. 
%SB E represent an extended torus surrounding the core (Maiolino et al.1997)

%Although our spectroscopic data are not good enough to decide for 
%sure, the fact that the major axis of the mini-spiral is well aligned 
%with the radio features suggests that the radio structure 
%is not an outflow. 

%%% CONCLUSIONS %%%%%
\section{Conclusions}
We have observed NGC\,3227 at $J$, $H$, and $K$ bands using adaptive optics on 
the CFHT. Despite artifacts surrounding the cores of bright
point sources with adaptive optics, we are confident that we have identified
several structures in the central few arcseconds.
An assembly of knot structures lying in a spiral/ring pattern within a
2\arcsec\ radius is 
suggestive of embedded spiral arms within the larger-scale spiral of the outer 
galaxy. 
With the recent discovery of spiral dust lanes in many Seyfert galaxies
observed with HST-NICMOS (Regan \& Mulchaey 1999), we suggest that our
observed spiral structure may be a fueling mechanism for the AGN.

An elongation of PA $\sim 40^\circ$ may be an unresolved continuation of the 
spiral pattern, or a nuclear disk. The coincidence of this feature within
the counter rotating molecular gas suggests the inner warped
disk proposed by Schinnerer et al.~(2000) as an interpretation.
%NOT A BAR too twisted 
The fact that the elongation aligns fairly well with a larger scale
[OIII] extended emission line region, and has rather blue colours, suggests
scattered nuclear continuum emission may also be responsible.
The extended radio feature observed with MERLIN is, however, not aligned
with the [OIII] axis, and is coincident 
with knot structures in the spiral/ring region. 

It is probably true that there must be a variety of mechanisms
responsible for ultimately getting gas into the central parsec of an
active galaxy.  Dense blobs spiraling in from 100's of parsecs due to
a viscosity provided by a hot outflowing may be one possibility,
though cloud stability is clearly a major problem in this case.
Cloud-cloud interactions may also play a role and effectively provide
a quasi-spherical cloud distribution in this region, providing a
natural setting for Malkan's galactic dust model (GDM) unification 
scenario.

There are still several conflicting pictures which appear consistent with the
data. 
The situation will be further elucidated by higher spatial resolution imaging
spectroscopy and perhaps imaging of even smaller-scale morphology with
adaptive optics on large telescopes.

%acknowledge
%
\section*{ACKNOWLEDGMENTS}

The research of SCC and GAHW was supported by a grant to GAHW from the
Canadian Natural Sciences and Engineering Research Council.
We would like to thank the staff at CFHT for facilitating
these observations. We acknowledge valuable comments from G.~Fahlman,
D.~Scott, M.~deRobertis, and an anonymous referee.

\end{document}